\documentclass[journal,onecolumn]{IEEEtran}
\usepackage{graphicx,subfig,amsthm,amsmath,latexsym,amssymb,times}
\usepackage{epsfig,multirow,rotating,times,verbatim,wrapfig,booktabs,float}
\usepackage{color,xr,array, algpseudocode}
\usepackage[compress]{cite}
\providecommand{\tightlist}{%
  \setlength{\itemsep}{0pt}\setlength{\parskip}{0pt}}

\usepackage{array}

\usepackage[T1]{fontenc}
\usepackage[utf8]{inputenc}
\usepackage{longtable}
\usepackage{pdfpages}
\usepackage{standalone}
\newtheorem{definition}{Definition}

\usepackage{placeins}

\newcommand{\bfM}{\mathbf{M}}

\newcommand{\citet}{\cite}
\newcommand{\citep}{\cite}

\newcommand{\bSigma}{\boldsymbol{\Sigma}}

\newcommand{\bPsi}{\boldsymbol{\Psi}}

\newcommand{\bOmega}{\boldsymbol{\Omega}}

\newcommand{\bff}{\boldsymbol{f}}

\newcommand{\bS}{\boldsymbol{S}}

\newcommand{\bx}{\boldsymbol{x}}
\newcommand{\bX}{\boldsymbol{X}}

\newcommand{\mN}{\mathcal N}

\newcommand{\bit}{\begin{itemize}}
\newcommand{\eit}{\end{itemize}}
\newcommand{\ben}{\begin{enumerate}}
\newcommand{\een}{\end{enumerate}}
\newcommand{\beqn}{\begin{equation}}
\newcommand{\eeqn}{\end{equation}}
\newcommand{\bea}{\begin{eqnarray*}}
\newcommand{\eea}{\end{eqnarray*}}
\newcommand{\bpf}{\begin{proof}}
\newcommand{\epf}{\end{proof}\ms}
\newcommand{\ms}{\medskip}

\begin{document}
\title{Fracture Mechanics-Based Quantitative Matching of Forensic  Evidence Fragments}

\author{Geoffrey~Z.~Thompson,~Bishoy~Dawood,~Tianyu~Yu,~Barbara~K.~Lograsso,~John~D.~Vanderkolk,\allowbreak~Ranjan~Maitra,~William~Q.~Meeker~and~Ashraf~F.~Bastawros
  \thanks{G. Z. Thompson is with the Department of Statistics, Indiana
    University, Bloomington, Indiana, USA.}
  \thanks{B. Dawood, T. Yu and A. F. Bastawros are with the Department
    of Aeronautical Engineering, B. K. Lograsso is with the Department
    of Mechanical Engineering, and W. Q. Meeker and R. Maitra arewith
    the Department of Statistics, all at Iowa State 
    University, Ames, Iowa, USA.}
  \thanks{John D. Vaderkolk is with Indiana State Police Laboratory,
    Fort Wayne, Indiana, USA.}
  \thanks{This research was supported in part by the
U.S. Department of Justice under its contracts
No.~2015-DN-BX-K056 and 2018-R2-CX-0034.
The content of this paper however is solely the responsibility of the 
authors and does not represent the official views of the NIJ.} 
}


\maketitle

\begin{abstract}
Fractured metal fragments with rough and irregular surfaces are often
found at crime scenes. Current forensic practice visually inspects the
complex jagged trajectory of fractured surfaces to recognize a ``match'' using comparative microscopy and physical pattern analysis.
We developed a novel computational framework, utilizing the basic concepts of fracture mechanics and statistical analysis to provide quantitative match analysis for match probability and  error rates.  The framework employs the statistics of fracture surfaces to become non-self-affine with unique roughness characteristics at relevant microscopic length scale, dictated by the intrinsic material resistance to fracture and its microstructure. At such a  scale, which was found to be greater than two grain-size or micro-feature-size, we establish that the material intrinsic properties, microstructure, and exposure history to external forces on an evidence fragment have the premise of uniqueness, which quantitatively describes the microscopic features on the fracture surface for forensic comparisons. The methodology utilizes 3D spectral analysis of overlapping topological images of the fracture surface  and classifies specimens with very high accuracy using statistical learning. Cross correlations of image-pairs in two frequency ranges are used to develop matrix variate statistical models for the distributions among matching and non-matching pairs of images, and provides a decision rule for identifying matches and determining error rates. A set of thirty eight different fracture surfaces of  steel articles were correctly classified. The framework lays the foundations for forensic applications with quantitative statistical comparison across a broad range of fractured materials with diverse textures and mechanical properties.

\end{abstract}



\subsection*{Introduction}

Consider the example of a crime scene where investigators found the tip of a knife or
other tool which broke off from the rest of the object. Later, investigators
recover a base which appears to match and they wish to show the two pieces are from
the same knife in order to use that evidence later at trial.
Scientific testimony used in a criminal or civil trial must be ``not only relevant but reliable'', according
to the Supreme Court decision Daubert v. Merrell Dow Pharmaceuticals, Inc (1993). The application
of this ruling forced a reconsideration of some previously acceptable forensic
evidence and a re-evaluation of the scientific validation of its premises
and techniques~\citep{fradella2003}. In 2009, The National Academy of Sciences (NAS)
issued a report, ``Strengthening Forensic Science in the United States: A Path Forward'',
which evaluated the state of forensic science and concluded that,
``[m]uch  forensic  evidence---including,  for  example,  bite  marks  and
firearm  and  toolmark
identification---is introduced in criminal trials without any meaningful scientific validation,
determination of error rates, or reliability testing to explain the limits of the
discipline.\cite{NRC_2009}''
The report highlighted the need to develop
new methods which have meaningful scientific validation and are accompanied
by statistical tools to determine error rates and the reliability of the methods.
To that end, the NAS has recently published reports on the state of fire
investigation~\cite{fire2017forensic}
and latent fingerprint examination~\cite{thompson2017forensic}.

Fracture matching is the forensic discipline of determining whether
two pieces came from the same fractured object. This  relies on the principle
that fracture surfaces are unique and that the individual characteristics
of the fracture process leave surface marks on both surfaces that can be
identified in order to match fragments to each other reliably. Current forensic practice for fracture matching visually inspects the complex jagged
trajectory of fracture surfaces to recognize a match using comparative
microscopy and tactile pattern analysis~\cite{vanderkolk2009forensic,crimesceneinvestigation2016}.
Previous research has supported that the observed
fracture patterns in metals are unique~\cite{Katterwe05,miller2006metal} and that
microscopic inspection of the fracture surfaces by examiners can reliably validate
matches~\cite{Claytor10}. 
However, this relies on subjective comparison without a statistical foundation,
which may be flawed:
``But even with more training and experience using newer techniques,
the decision of the toolmark examiner remains a subjective decision based on
unarticulated standards and no statistical foundation for estimation of error
rates.~\cite{NRC_2009})''
It is therefore desirable to develop more objective methods using quantitative measures
that can be validated with less human input for use in a criminal or civil trial.

\begin{figure}[ht]
  \centering
  \includegraphics[width=\textwidth]{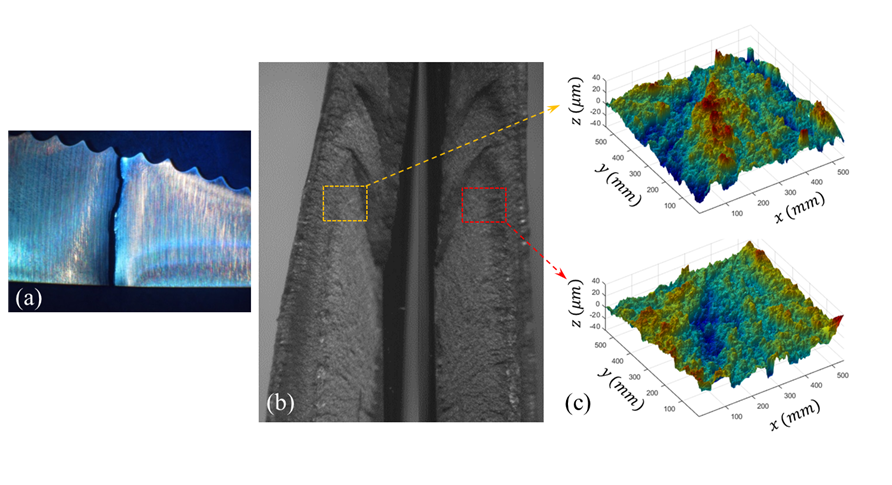}
  \caption{Association of forensic fragments. (a) Visual jigsaw match of the macroscopic crack trajectory. (b) Physical pattern match with comparative microscopy. (c) 3D representation of fracture surface, showing detailed topographic features at relevant scale. \label{fig:fig1} }
\end{figure}

Here we propose a statistical method guided by the physics of fracture mechanics
to perform forensic fracture matching using imaging of microscopic fracture
details. The basis for physical matching is the assumption that there is an
indefinite number of matches all along the fracture surface. The
irregularities of the fractured surfaces are considered to be unique and
may be exploited to individualize or distinguish correlated pairs of
fractured surfaces~\cite{vanderkolk2009forensic,van2004physical}. For example, the complex jagged trajectory of
a macro-crack in the forensic evidence specimen of Figure~\ref{fig:fig1}(a) 
can sometimes
be used to recognize a ``match'' by an examiner or even by a layperson on a jury~\cite{vanderkolk2009forensic, van2004physical}. However, experience, understanding, and judgment are
needed by a forensic expert, to make reliable examination decisions
using comparative microscopy and physical pattern match as indicated in
Figure~\ref{fig:fig1}(b). Indeed, the microscopic details of the non-contiguous crack
edges on the observation surface of Figure~\ref{fig:fig1}(a, b) cannot always be
directly linked to a pair of fractured surfaces, except possibly by a
highly experienced examiner. There are many published studies and case
reports concerning fracture matching of different materials such as
rubber shoe soles, wood, glass, tape, paper, skin, fishing line, cable,
and, most commonly, metal~\cite{Klein00,Walsh94,Matricardi75,McKinstry98,
  verbeke75,Townshend76,Dillon76,Karim04,Smith04,Katterwe83,Goebel83,Moran84,
  Rawls88,Hathaway94,Zheng_2014,petraco12_addres_nation_academ_scien_chall}.
However, the microscopic details
imprinted on the topological fracture surface of Figure~\ref{fig:fig1}(c) carry
considerable information that could provide a quantitative forensic
comparison with higher evidentiary value. Forensically, glass and metal
fracture surfaces were shown to have highly stochastic fracture-branches due to the randomness of the
microstructure and grain sizes~\cite{Katterwe05,katterwe1982comparison}, with limited prior attempts to
quantitatively match two measured fracture surface topologies~\cite{Matricardi75,Claytor10}.

The rough and irregular metallic fracture surfaces carry many details of
the metal microstructure and its loading history. Mandelbrot et
al.~\cite{Mandelbrot84} first showed the self-affine scaling properties of fractured
surfaces to quantitatively correlate the material resistance to fracture
with the resulting surface roughness. The self-affine nature of the
fracture surface roughness has been experimentally verified for a wide
range of materials and loading conditions. A key finding is the
variation of such surface descriptors when measured parallel to the
crack front and along the direction of propagation \cite{Ponson07,Alava2006,Bonamy11, Yavas2021}. Additionally,
while self-affine characterization of the crack
surface roughness exists at a length scale smaller than the fracture
process scale (where stresses ahead of the crack tip reach critical value) ~\cite{Anderson2017fracture}, the surface character becomes more complex and
non-self-affine at larger length scales~\cite{Cherepanov95}.

We first present an overview of the method and the study objectives. Then we describe the sample
generation method and the imaging process used to create the data. We then provide a description of the statistical model
which discriminates the matching fracture surfaces from the non-matching surfaces.
We provide an evaluation of the method and several experiments to guide choices
in imaging and in the parameters for the statistical model. Finally, we provide a discussion
of the results and an illustration of how it would be applied in a forensic context.
In the supplementary materials, we provide the underlying data, code to reproduce the analysis and figures, and additional information about the methods and materials. An R software package to perform the model fitting
and analysis, \texttt{MixMatrix}, is available~\cite{thompsonmix}.


\subsection*{Method Overview and Study Objectives}\label{study-objective}
\addcontentsline{toc}{subsection}{Study Objectives}

Our objective is to find the scale of unique features on
a fracture surface and then create a statistical method which uses
the features to match them in a way which
is suitable for use as evidence in court.
Motivated by the observations about the self-affine nature of fracture surfaces, it can be speculated that a randomly
propagating crack will exhibit unique fracture surface topological
details when observed from a global coordinate that does not recognize
the direction of crack propagation. This work explores the existence of
such uniqueness of a randomly generated fracture surface at some
relevant length scales. The uniqueness of these topological features
implies that they can be used to individualize and distinguish the
association of paired fracture surfaces. Our approach uses the fact that the
microscopic features of the fracture surface in Figure~\ref{fig:fig1}(c) possess unique
attributes at some relevant length scale that arise from the interaction
of the propagating crack-tip process-zone and the microstructure details.
The corresponding surface roughness analysis of this surface is shown in
\begin{figure}[ht]
  \centering
  \includegraphics[width=\textwidth]{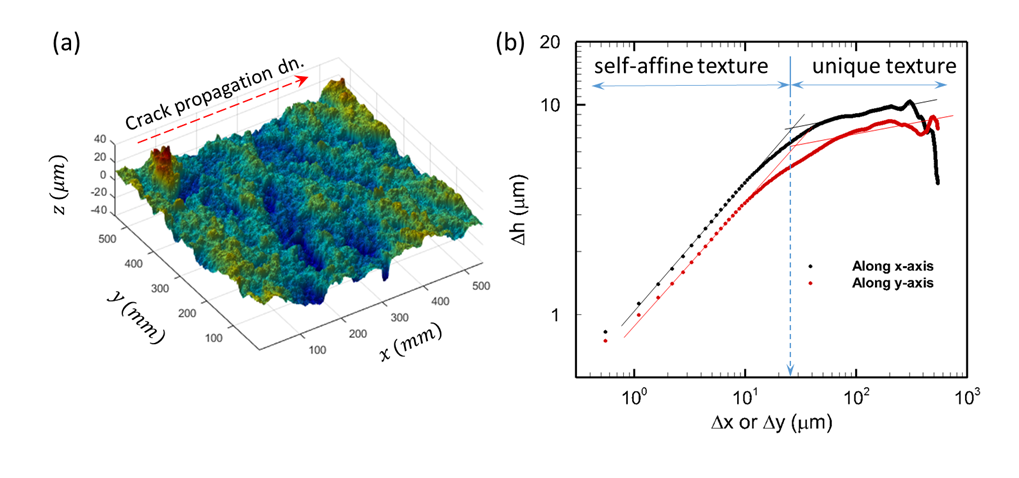}
  \caption{Fracture surface characteristics. (a) 3D surface topology rendering of fractured surface, showing a biased orientation of the low-frequency texture of the fracture surface. The direction of crack propagation is along the $x$-axis (b) Height-height correlation variation with the size of the imaging window, showing the domain of the self-affine deformation and the deviation of the fracture surface characteristics at higher length scales ($ > 100\mu m$), which could be used for matching purposes.      \label{fig:fig2} }
\end{figure}
Figure~\ref{fig:fig2} using a height-height correlation function, 
$\delta h(\delta \mathbf{x}) =  \sqrt{\langle[h(\mathbf{x} + \delta \mathbf{x})-h(\mathbf{x})]^{2}\rangle_{\mathbf{x}}}$, 
where $\langle \rangle$ denotes averaging over the $x$-direction. We see that at the
small length scale of less than 10 -- 20 $\mu m$, the roughness
characteristic is self-affine (i.e. proportional to the analysis window scale). However, at larger length scales
($\approx 100 \mu m$), this characteristic deviates, showing the
individuality of the surface at that scale. These microscopic feature signatures 
exist on the entire fracture surface as it is influenced by three
primary factors; namely the material microstructure, the intrinsic
material resistance to fracture, and the direction of the applied load. This
work explores the existence of such a length scale and the corresponding
unique attributes of the fracture surface, as well as their applications to forensic comparison of fractured surfaces.

The height-height correlation function at this transition scale captures the uniqueness of
the fracture surfaces, so we can use that function's behavior in setting the observation scales for comparing matching and non-matching
surfaces to produce a statistical model of each topological class's behavior for use in classification.
We can further combine multiple observations at different length scales or topological frequencies of a single surface into one model in order to
improve the ability to discriminate between surfaces of the same class or materials and manufacturing processes (for instance, individualization of a pry tool from a similar batch of identical tools).
The statistical model can produce a likelihood ratio or log-odds ratio of a new set of surfaces
belonging to either class, which are common outputs of forensic matching methods~\cite{aitken04_statis_evaluat_eviden_foren_scien,
  meester04_why_effec_prior_odds_shoul,keijser12_under_foren_exper_repor_by,
  martire14_inter_likel_ratios_foren_scien_eviden,
  zadora13_statis_analy_foren_scien,taroni14_bayes_networ_probab_infer_decis}.
The creation of this model can also be used to estimate probabilities of misclassification
and compare to the empirically observed rates of misclassification.
Conceptually, this is similar to forensic matching models which are used in fingerprint identification and
bullet matching. In fingerprint identification, features (minutiae) on the reference print
and the latent print are marked and then the pair is given a score based on how well the
two match, which may be part of a probabilistic model reporting a likelihood ratio or
other probabilistic output~\cite{Champod2016}. The Congruent Matching Cells approach for
matching breech face impressions on cartridge cases in ballistics takes a similar approach:
it divides the scanned surfaces into cells and searches for matching cells on the other
surface. It then uses this as an input to a statistical model which outputs a likelihood
ratio~\cite{Song_2013, Chen2018}.


\subsection*{Materials and Methods}\label{materials}
\addcontentsline{toc}{subsection}{Materials and Methods}

\subsubsection*{Sample Generation and Imaging}\label{sample}
\addcontentsline{toc}{subsubsection}{Sample Generation and Imaging}

We consider two main material classes: sets of rectangular rods of a
common tool steel material (SS-440C) and sets of knives from the same
manufacturer fractured under control tension and bending configurations, respectively. 
The average grain size for both groups was approximately $dg$ = 25--35 $\mu m$.
Four different sets of samples were established with nine specimens in the two sets of knives
and ten specimens in the two sets of steel rods. Each
knife specimen was fractured at random, in a manner similar to Figure~\ref{fig:fig1}(a).

For clarity, we refer to the surface attached to the knife handle as the
base and the surface from the tip portion of the knife as the tip and apply the same terminology to samples from the rectangular steel rods. The microscopic
features of pairs of fracture surfaces were analyzed by a standard
non-contact 3D optical interferometer (Zygo-NewView 6300), which
provides a height resolution of 20 $nm$ and spatial inter-point resolution
of $0.45\mu m$ (Figure~\ref{fig:fig2}(a)) at an optical magnification of 20X. Surface height 3D topographic maps were
acquired from the pairs of fracture surfaces, and quantized using
Fourier transform based spectral analysis as summarized in
Figure~\ref{fig:fig3} in the image analysis step. Further details are given in
the Supplement. 
The unique implementation of frequency space analysis provides a greater
tolerance for the alignment of the pair of images.
Further, it provides a straightforward segmentation of the surface
topological frequency ranges for comparison.  

The analysis first identified the scale of the significant features on each
image pair and their distributions. It is established in fracture
mechanics that the fracture process zone ahead of the crack tip
typically extends to 2-3 times the grain size ~\cite{Anderson2017fracture}, or around 50--75 $\mu m$
for the tested material system. This is the scale wherein the local stresses ahead of the crack tip reach a critical level, sufficient to overcome the intrinsic resistance of the material to fracture ~\cite{Anderson2017fracture}.  Typically, a field of view (FOV) that covers
at least 10 periods of the fracture process zone or about 20-30 grain
diameters should be utilized to avert signal aliasing. An extended set
of nine topological images with a $550 \mu m$ FOV was
collected on each fracture surface.

\begin{figure*}[ht]
  \centering
  \includegraphics[width=\textwidth]{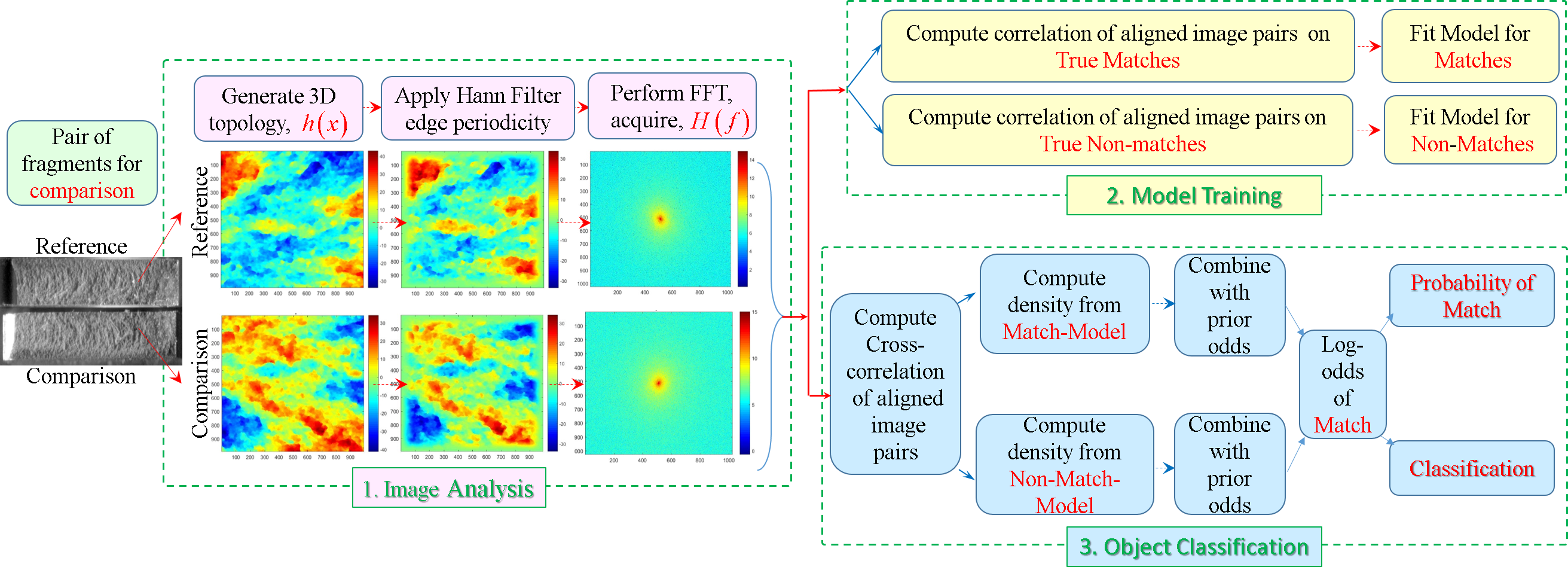}
  \caption{Flow chart showing the image analysis steps and model fitting/calibration step, followed by 
  classification of new objects to provide classification probabilities. For a new field object, an examiner would use Step 1 to image the object and perform Step 3 using a model trained in Step 2 on samples of the same class to guide forensic conclusions.\label{fig:fig3} }
\end{figure*}

\subsubsection*{Physical Matching by Spectral Analysis and Image Correlation}\label{physical-matching}
\addcontentsline{toc}{subsubsection}{Physical Matching by Spectral
Analysis and Image Correlation}

The measured height distribution function $h(\bx)$, defines the topology
of the fracture surface at every spatial point, $\bx$ on the fracture
surface. 
Each wavelength on the
fracture surface has a population,  on the frequency domain $H(\bff)$, which
is acquired using a Fast Fourier Transform (FFT) operator. For example, grain size
has a distribution of frequencies across the spectrum rather than one specific frequency.
Similarly, other microscopic fracture features would have a
range of spectral distributions~\cite{kobayashi10_fract_surfac_topog_analy_frast,Jacobs_2017}.
For a pair of fractured surfaces, the
population of these features contains relevant information about the
physical processes present at each length-scale. After calculating the
spectra of each pair of images, each spectrum was divided into multiple
radial sectors. The segmented angular sectors for the frequency range
($0^\circ$, $180^\circ$) represents the entire data set, because
the amplitude of $H(\bff)$ exhibits inversion symmetry. The
spectral array size is proportional to $2^n$, as this is a mathematical
feature of the FFT. For the image size employed in this work,  a spectral array of 1024 by 1024 is acquired, although only the upper half is utilized because of symmetry. The radial segments for
comparison on the frequency domain are chosen to reflect the physical process scales and the
corresponding wavelength.

\begin{figure}[ht]
  \centering
  \includegraphics[width=\textwidth]{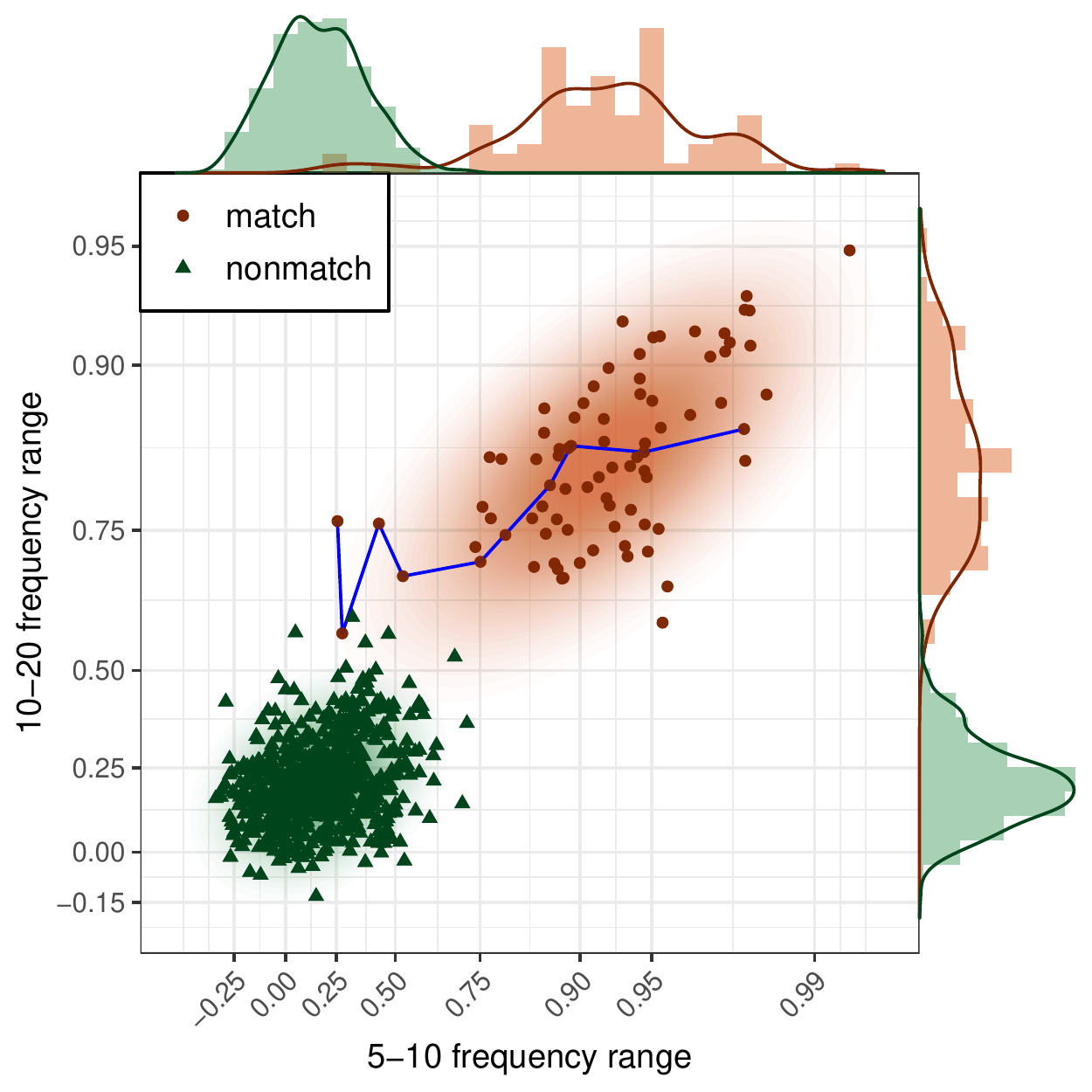}
  \caption{Scatter plot of correlations for 81 matched pairs and 648 non-matched pairs for the 5--10 and 10--20 $mm^{-1}$ frequency ranges on a Fisher-z (nonlinear) axis. The plot  shows that true matches and true non-matches
    are distinguished in this example by features in the 5-10 and 10-20 $mm^{-1}$
    frequency ranges. The connected points show the values of nine images from the same surface, indicating that while some individual images may not distinguish matches from  non-matches, taking an ensemble of images from the surface helps perfectly discriminate the two classes. \label{fig:scatter} }
\end{figure}

For comparison, we use the frequency amplitude, $\bar H (\bff)$ for each surface spectral frequency.
To compare two surfaces, two-dimensional statistical correlations
between their spectra are computed in banded radial frequencies, with
increments in the bands determined by the scale of the image and the material microstructure, yielding a similarity measure on each frequency band for the corresponding
pairs of images.  To estimate the distribution for
both the population of true matches and true non-matches, this is done for images from matching fracture surfaces
and non-matching fracture surfaces  as shown in Figure~\ref{fig:scatter}. On every
fracture surface, a series of up to $k$-overlapping images were
collected for the comparison process and the establishment of a
statistical match. We used $k=9$ images with 75\%
overlap between successive images. The choice of overlap means
there are three full independent sequential images on a surface. 

The classification and matching process strategy is summarized in  Figure~\ref{fig:fig3}, and is carried out in two steps; (a) Model training on an initial data set and (b)
performing classification of new sample(s). 

\paragraph{(a) Model Training/Fitting:}\label{model-fitting}
\addcontentsline{toc}{paragraph}{(a) Model Training/Fitting:}

After determining which frequency bands are relevant for the comparison
of fracture surfaces in a given material class, a model to distinguish matching
from non-matching fracture surfaces can be developed. The behavior of
the frequency band correlations in the population of matches and
non-matches has to be estimated and modeled. The proposed framework  provides a separate model for each class
(i.e.~match and non-match). The model training process is highlighted in Figure~\ref{fig:fig3} and entails:

\begin{enumerate}
\def\labelenumi{\arabic{enumi}.}
\tightlist
\item
 Choose a set of experimental fracture pairs to train the  model.
\item
  Compute the correlations of the frequency bands for the sets of images for all matching and non-matching surface pairs.
\item
  Use the Fisher's $z$ transformation on the correlation data to stabilize variance~\cite{fisher1915frequency}.
\item
  Fit models to describe the distribution of true matches and true
  non-matches, which account for the difference in location of the
  correlations and account for the covariance of the repeated
  observations across the surface.
\end{enumerate}

\begin{figure}[ht]
  \centering
  \includegraphics[width=\textwidth]{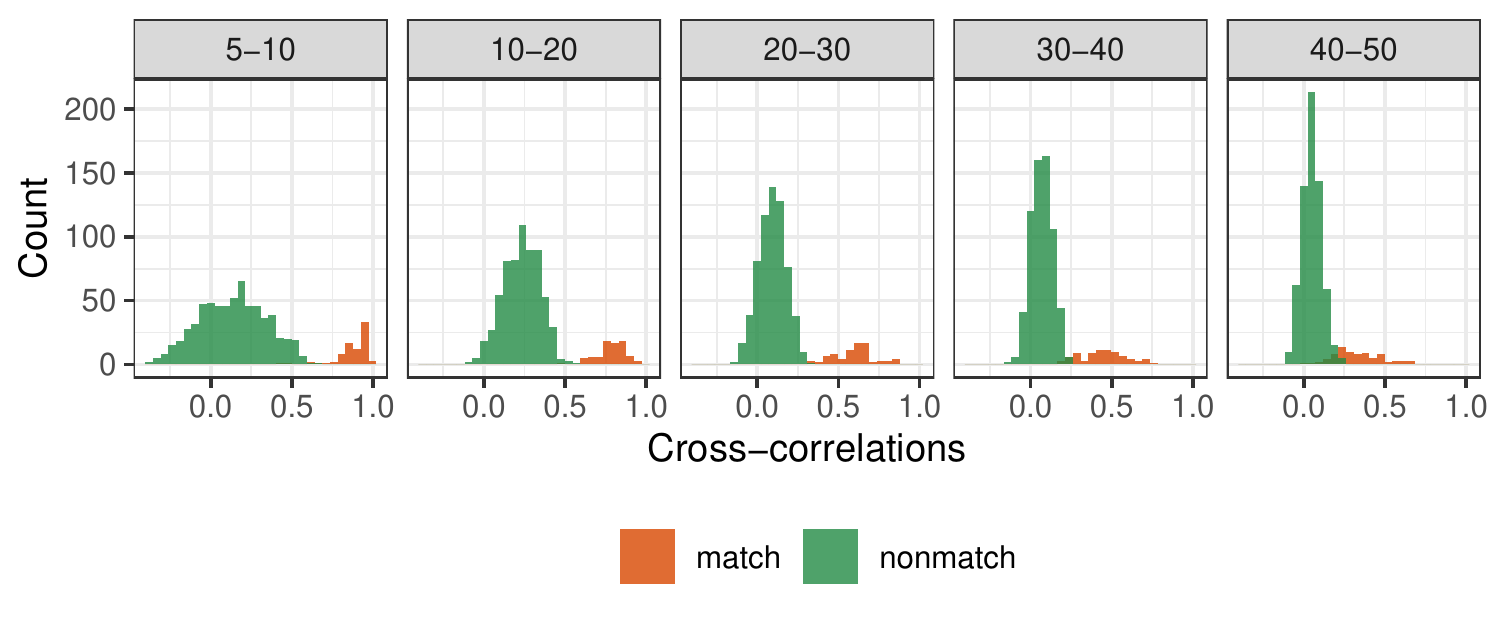}
  \caption{Histograms of correlations of true matches and true non-matches in
    one set of 9 surfaces with 9 images per surface split by frequency band. Lower
    frequencies are well-separated, but higher frequencies begin to have more substantial
    overlap.\label{fig:histograms} }
\end{figure}

Figure~\ref{fig:scatter} illustrates the discrimination ability of the proposed method.
The data in this illustration were derived from nine base-tip pairs from fractured knives.
Nine images were taken from each base and tip fracture surface,
resulting in 162 total images (81 from the tips and 81 from the bases).
In this example, image
pairs for when the tip and base surfaces were from the same knife are
true matches (81 matched-pairs), while those pairs for which the tip and base surfaces were
from different knives are true non-matches (648 unmatched-pairs). Correlation analysis showed
clear separation (lower values for the true non-matches
and higher values for the true matches) for the 5--10 and
10--20$mm^{-1}$ frequency-band range, as shown in Figure~\ref{fig:histograms}.
At the lower frequency ranges, there is some overlap. 
Beyond these frequency ranges, the
true match and the true non-match correlation distributions become less distinct and overlap more.
In the
set displayed in Figure~\ref{fig:scatter}, there is one image pair among the true matches which
cannot be distinguished from the true non-matches and three other pairs
that are ambiguous. To further improve the discrimination, considering multiple observations
from the same surface would distinguish it from the non-matches, since
the other observations on that surface are well-separated from the non-matches.

Because we have nine overlapping images for each specimen and two (or more) comparison frequencies,
each comparison between two specimens provides a $2 \times 9$ matrix of correlations.
Accordingly, we propose using a matrix-variate
distribution~\cite{gupta2018matrix, iranmanesh2010conditional} to model the densities of the matching and non-matching populations,
and, specifically, a matrix-variate  $t$ distribution (MxV$t$) because the data for the individual comparisons
are approximately elliptically
distributed but have heavier tails than a normal distribution. A definition
of the distribution is in the supplement and the density is defined in Equation S-1 in the Supplement.

We use matrix-variate distributions to model the relationship between the two frequency bands
in each image comparison and across the images being compared in the base and tip pair (e.g. Figure~\ref{fig:scatter}).
Because of the overlapping-image structure of the data source,
our model allows between-image correlations in the
matrix-variate model to be related according to an autoregressive
model of order 1 (or AR(1)) model (implying that immediately
adjacent images can be correlated).
We specify that
the mean correlations in the two frequency bands remain the same across the images
on a surface in the model.
The fit of the model is estimated using an expectation-maximization (EM) algorithm developed for the
matrix-variate $t$ distribution~\cite{thompson2020}.

\paragraph{Classification of a new object}\label{classification}
\addcontentsline{toc}{paragraph}{Classification of a new object}

Suppose the fitted model has been trained on a set of $k$-images
per fracture surface, yielding probability density functions $f_1$
corresponding to the population of true matches and $f_2$ corresponding to the
population of true non-matches. Suppose also that there is a new  pair of fracture surfaces
that may or may not match.
First, the correlations
for the $k$-aligned image pairs in the chosen frequency bands are
computed and transformed, yielding a new observation $X$, which is a matrix of
observations of correlations with one row for each frequency band and one column
for each pair of images---here, a $2 \times k$ matrix.
Then, presuming prior probability $p$ of
being a true match and prior probability $1 - p$ of being a
true non-match, we can find, by combining prior probabilities and the
match and non-match densities from the model, the posterior probability that the two
surfaces match as follows:

\[P(X = match) = \frac{p f_1(X)}{p f_1(X) + (1-p) f_2(X)}\]

In the absence of prior information of the probability of a match, we are using an equal prior ($p = 0.5$).
A classification decision can then be made based on the posterior probability. 
The results can be expressed as a log-odds ratio.
Changing the prior probabilities changes
the log-odds ratio by adding a constant, so the specification of a prior at this point is unnecessary. 
If an equal prior is used, this expression can also be
converted to a likelihood ratio (LR), which is a common method in
forensic applications~\cite{aitken04_statis_evaluat_eviden_foren_scien,
  meester04_why_effec_prior_odds_shoul,keijser12_under_foren_exper_repor_by,
  martire14_inter_likel_ratios_foren_scien_eviden,
  zadora13_statis_analy_foren_scien,taroni14_bayes_networ_probab_infer_decis},
and these LR results can be incorporated
into a framework for evaluating the strength of evidence under different
sets of assumptions ~\cite{lund17_likel_ratio_as_weigh_foren_eviden}.
Classification decisions can then be made under the
rules of evidence relevant to the case. In this discussion, we will
make classification decisions using a cutoff value of $0.5$.


\subsection*{Results and Discussion}\label{results}
\addcontentsline{toc}{subsection}{Results and Discussion}

\paragraph{Classification performance}\label{performance}
\addcontentsline{toc}{paragraph}{Classification performance}

\begin{figure}[ht]
  \centering
  \includegraphics[width=\textwidth]{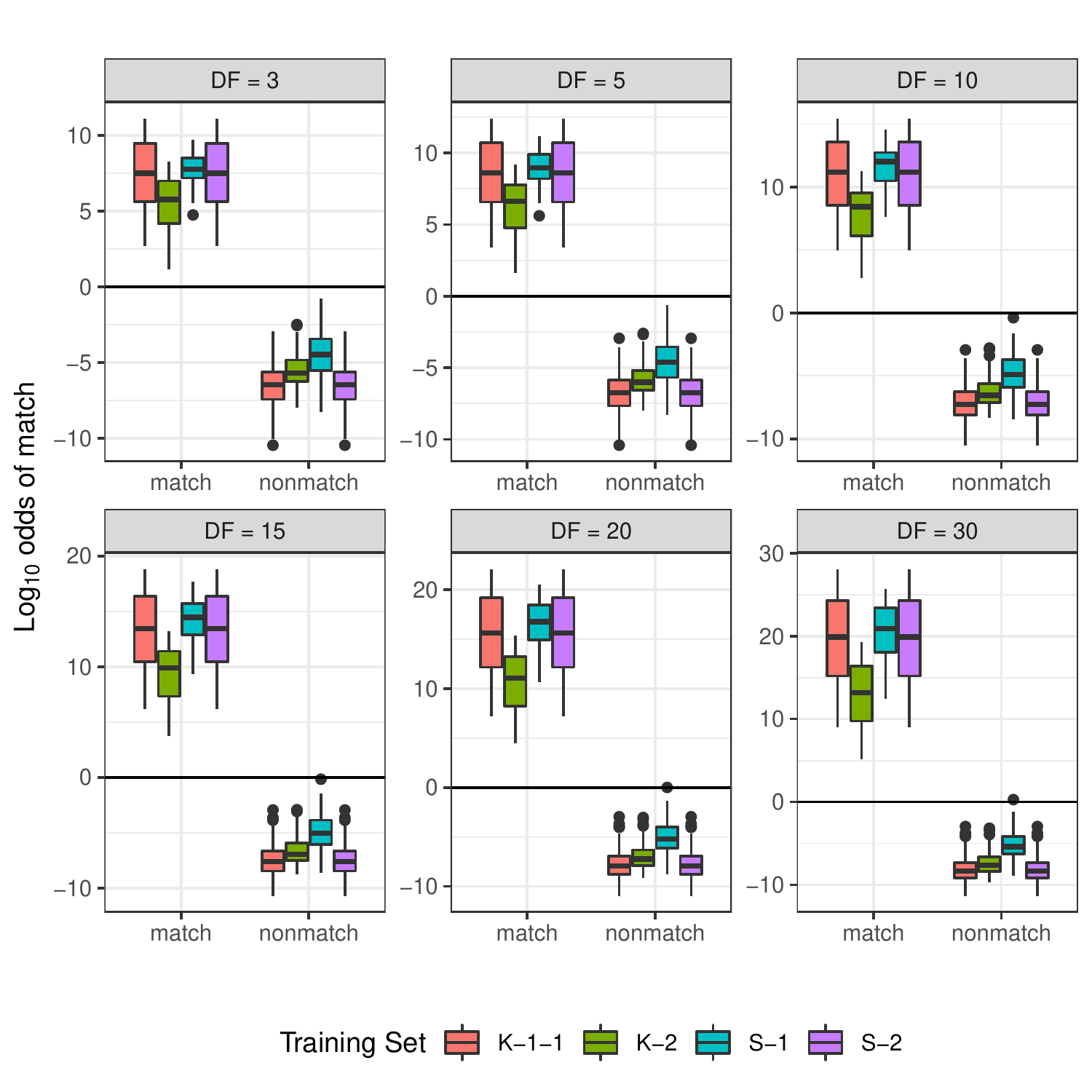}
  \caption{ Log-odds of being a match split by training set and true class membership for matrix $t$ distributions with 3, 5, 10, 15, 20, and 30 degrees of freedom. Log-odds greater than 0 indicates greater odds of being a match than a non-match. The predictions for each training set are on all four sets of surfaces.\label{fig:classification} }
\end{figure}

There are
two datasets from the knives and two from the steel bars:
``K-1-1'' is the first set of images from the first set of knives, and the imaging is independently repeated generating additional sets of images ``K-1-2''and ``K-1-3'' for repeat analysis. ``K-2'' indicates the other set of knives, whereas ``S-1'' and ``S-2'' indicate the two steel bar samples.
Figure~\ref{fig:classification} shows the classifications obtained by
training on each of the four datasets, represented by one of the color boxes, with all 9
images per sample and classifying
on all the sets of surfaces using the matrix-variate $t$ distribution and
a common degrees of freedom parameter, $\nu = 3, 5, 10, 15, 20$, and $30$.
The output given in terms
of the log-odds of being a match -- log-odds larger than zero $(p=0.5)$ indicate
classification as a match. While initially there are no false positives or false negatives, as the degrees of freedom
parameter (DF or $\nu$) increases, there is one false positive, though this probability is very close to 0.5 and
all of the true positives have a probability close to 1, which suggests using a
classification threshold other than 0.5 would yield perfect classification.
A different threshold can be chosen by selecting a low probability (such as $10^{-4}$)
as a probability of false alarm and using the distribution of log-odds of the true non-matches
to fix that threshold conservatively by selecting an upper confidence bound of that quantile~\cite{meeker2017statistical}.
Using the upper $95\%$ confidence bound for the threshold at which the false alarm probability based
on the distribution of true negatives is  $10^{-4}$ sets the threshold at a probability of 0.8375
for the most conservative training set at the setting of $\nu = 10$, for example, which still results
in perfect classification.

\begin{figure}[ht]
  \centering
  \includegraphics[width=\textwidth]{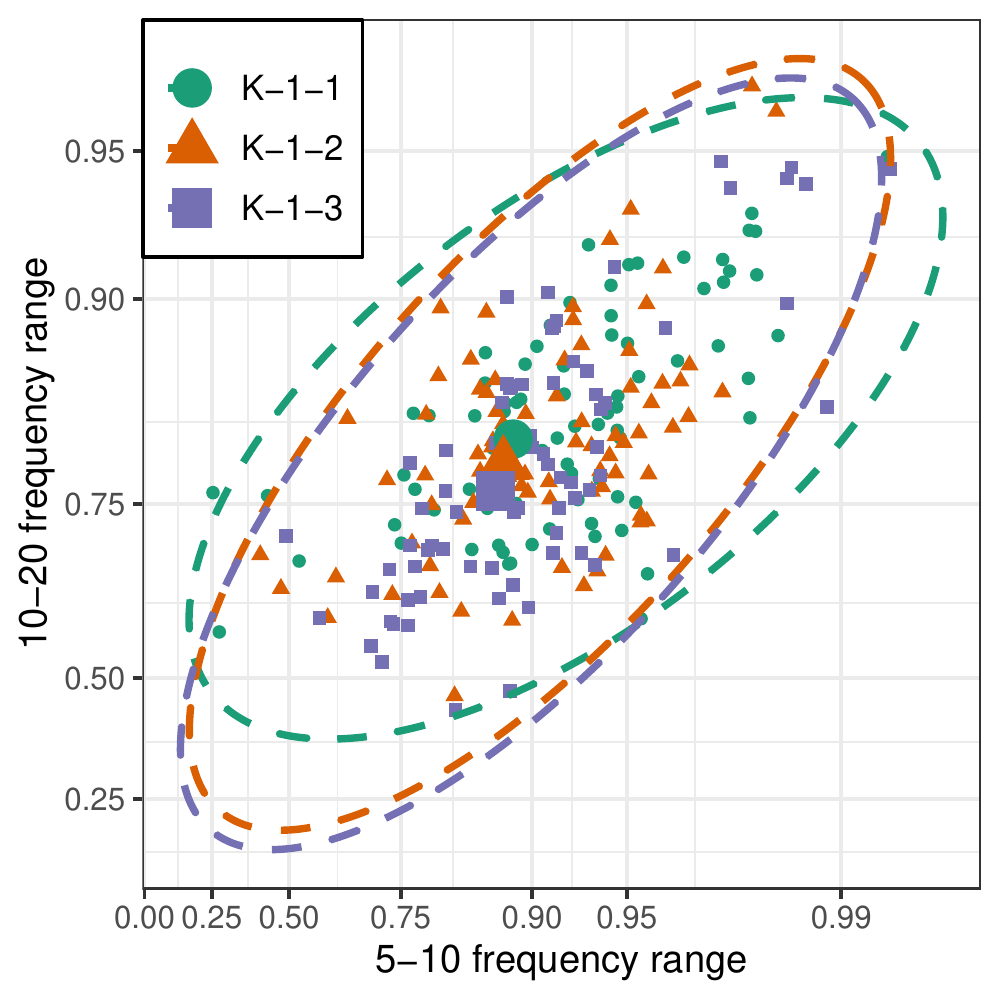}
  \caption{Individual true match correlations for three repetitions of images
    of the same set of 9 knives with 9 images per knife. This demonstrates
    that similar results will be obtained upon re-imaging the same surface,
    which is important in forensic applications. The large dots indicate the means of the sets and the ellipses are 99\% tolerance ellipses, which provide a representation of the covariance matrices.\label{fig:reproducibility} }
\end{figure}

\paragraph{Reproducibility of results}\label{reproducibility}
\addcontentsline{toc}{paragraph}{Reproducibility of results}
In order to determine the reproducibility of results for
a given sample, we re-imaged one of the knife samples three times
and examined the distributions of the true match image correlations in Figure~\ref{fig:reproducibility}. The different re-imaged sets are
labeled ``K-1-1'', ``K-1-2'', and ``K-1-3''. The means of the
distributions (indicated by the large shapes) are similar and the covariance matrices,
visualized using 99\% confidence ellipses, are also similar. Using the
two-sample Peacock test, a two-dimensional extension of the Kolmogorov-Smirnov
test~\cite{peacock83_two_dimen_goodn_of_fit, xiao17_fast_algor_two_dimen_kolmog},
there is no evidence these distributions differ
($p$-values: 1 and 2, $p=0.21$; 1 and 3, $p=0.32$; 2 and 3, $p=0.25$). We conclude, then, the imaging and analysis process are reproducible for the analyzed samples. 

\paragraph{Selecting DF ($\nu$)}\label{numberdf}
\addcontentsline{toc}{paragraph}{Selecting DF ($\nu$)}

\begin{figure}[ht]
  \centering
  \includegraphics[width=\textwidth]{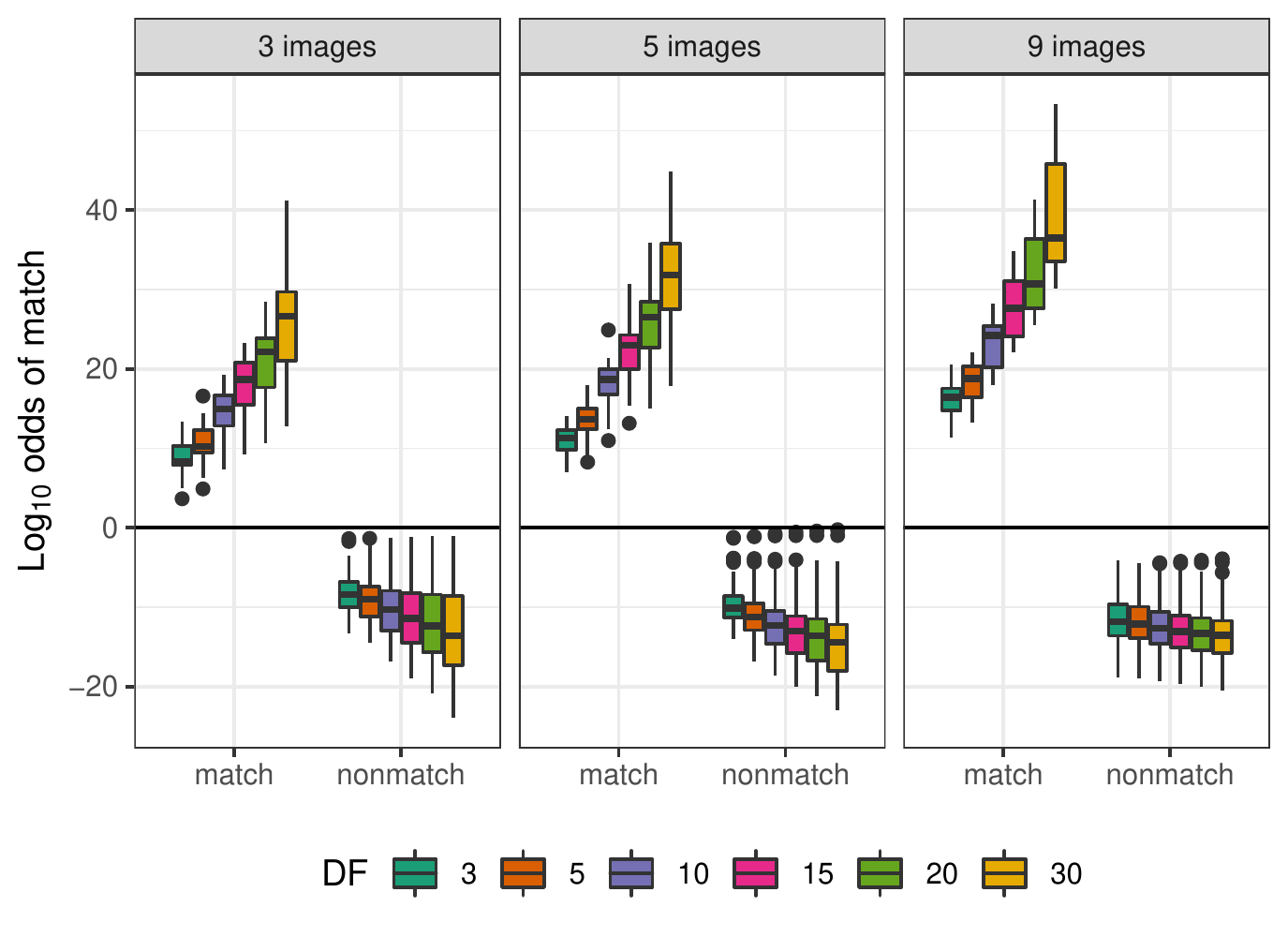}
  \caption{Cross-validation results for models fit using $k=3$, $5$, and $9$ images of each surface. The cross-validation was done to provide guidance about the number of images and the choice of DF ($\nu$). There were no false positives or false negatives in this analysis, so it did not provide any conclusive results. \label{fig:cvnine} }
\end{figure}

The training sets do not have a sufficient number of observations in both classes to estimate
$\nu$ in the MxV$t$ model. However, the analysis in the previous section
indicates it has some influence on the results. We performed a leave-one-out
cross validation (LOOCV) procedure to provide guidance about the effects of changing the parameter.
For each surface in a training set, a model was trained on the set of observations excluding that surface
and tested on the observations using the excluded surface. This was done for $k=9$ images on training sets S-1 and S-2 and
using $k = 5$ images (restricting to the images with only 50\% overlap) and $k = 3$ images (restricting
to the non-overlapping images) on all four training sets.
The procedure was performed only on sets S-1 and S-2 for $k=9$ because nine surfaces are needed to fit
the model and K-1-1 and K-2 have only nine surfaces, while S-1 and S-2 have ten. Figure~\ref{fig:cvnine} shows the results for $k=3$, $5$,  and $9$ respectively. The parameter $\nu$ varied from $3$ to $30$. In all cases, the true matches and true non-matches were
perfectly classified using a threshold probability of $0.5$ (log-odds of $0$).
Higher values of $\nu$ had more separation between the classes. Using 9 images
with 75\% overlap had greater separation than 5 images with 50\% overlap and greater separation between
the identification of true matches. However, given that there is perfect classification in all cases,
this does not provide much guidance on the selection of $\nu$.

\paragraph{Number of images needed for discrimination and model selection}\label{number}
\addcontentsline{toc}{paragraph}{Number of images needed}
Due to unique topological disturbance in some images (e.g. grains fall out from the fracture surface or significantly large out of plane curvature within the range of comparisons), there is not perfect separation between all image pairs for the matches
and non-matches. This can be noticed on Figure~\ref{fig:reproducibility} where some image pairs have a correlation coefficient of less than 0.50 for the two bands of frequency analysis.  To mitigate the influence of local topological disturbances when
deciding whether a pair of fragments represent a match or not,
multiple observations are needed.
To determine how many
images are needed to optimize classification performance, we started by training
models using all nine images on each training set as before. 
We again used the MxV$t$ model with $\nu = 3, 5, 10, 15, 20$,
and $30$,

and then tested them on 
subsets of consecutive images of size $k$, for $k = 2, 3, \ldots 9$
with the model reduced to considering only the selected images.
A summary of the complete results are given in the Supplement.

In Figure~\ref{fig:fpforallsubsets}, models with higher $\nu$ have higher false negative rates for all
values of $k$. For values of $k$ over 4, only $20$ and $30$ DF have false negatives.
Low values of the degrees of freedom parameter have false positives. All of
this suggests that choosing a value near $\nu = 10$ and $k \geq 5$ images are sufficient
for optimal classification. Figure~\ref{fig:allsubsets} displays complete results
for a model with 10 DF. As $k$ increases,
the typical classification results become more separated. However, even with only
two images considered in the test cases for the 10 DF model, the accuracy is very high and the worst case
of a false positive is classified with only a probability of 0.921 and the worst case of a false
negative is classified with a probability of 0.504. 
\begin{figure}[ht]
  \centering
  \includegraphics[width=\textwidth]{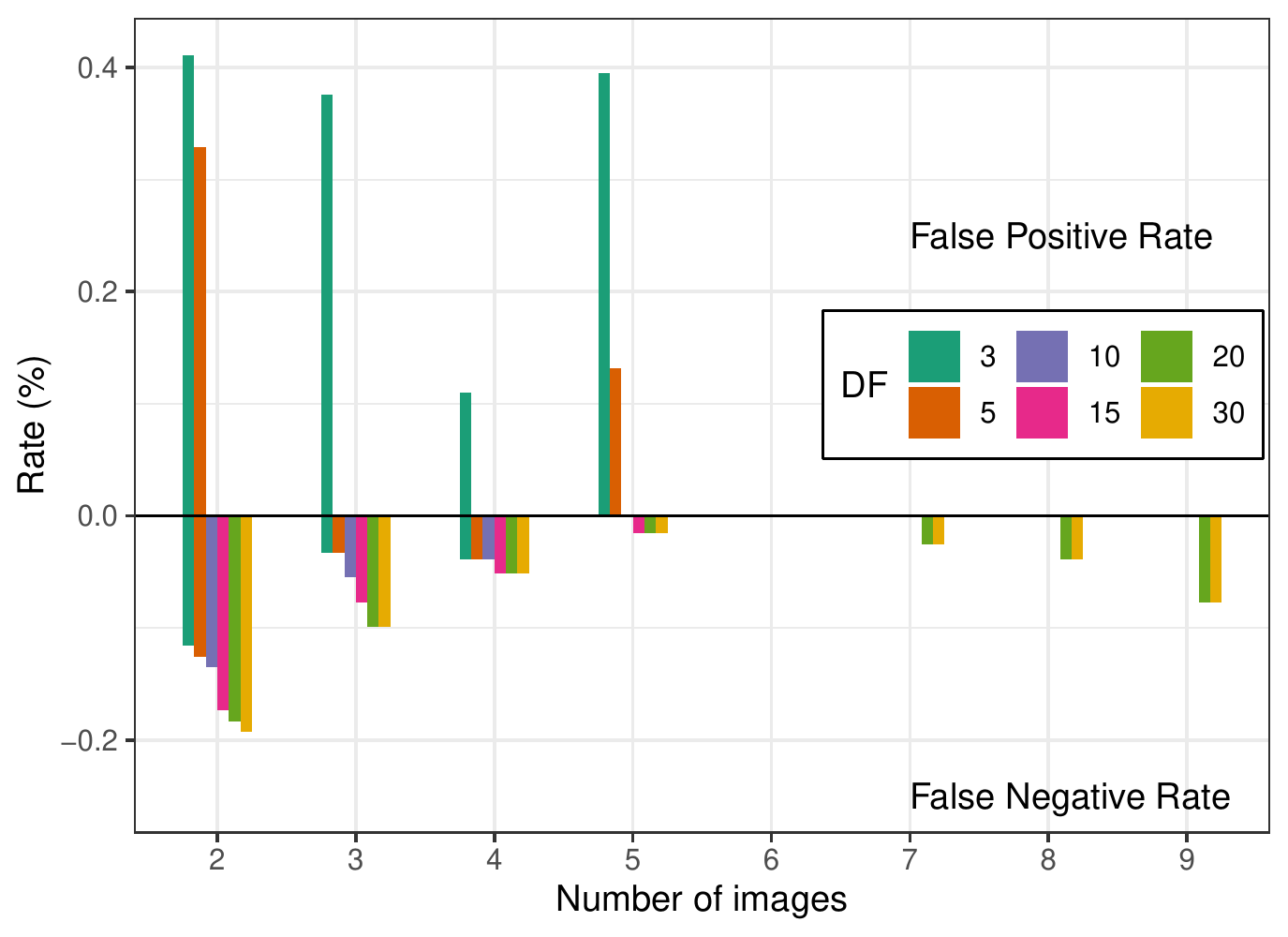}
  \caption{Rates of false positive and false negative classifications (in \%) using models trained
    on the four different sets of surfaces and tested on consecutive subsets
    of those images for $k = 2, 3, \ldots, 9$. \label{fig:fpforallsubsets}}
\end{figure}

\begin{figure}[ht]
  \centering
  \includegraphics[width=\textwidth]{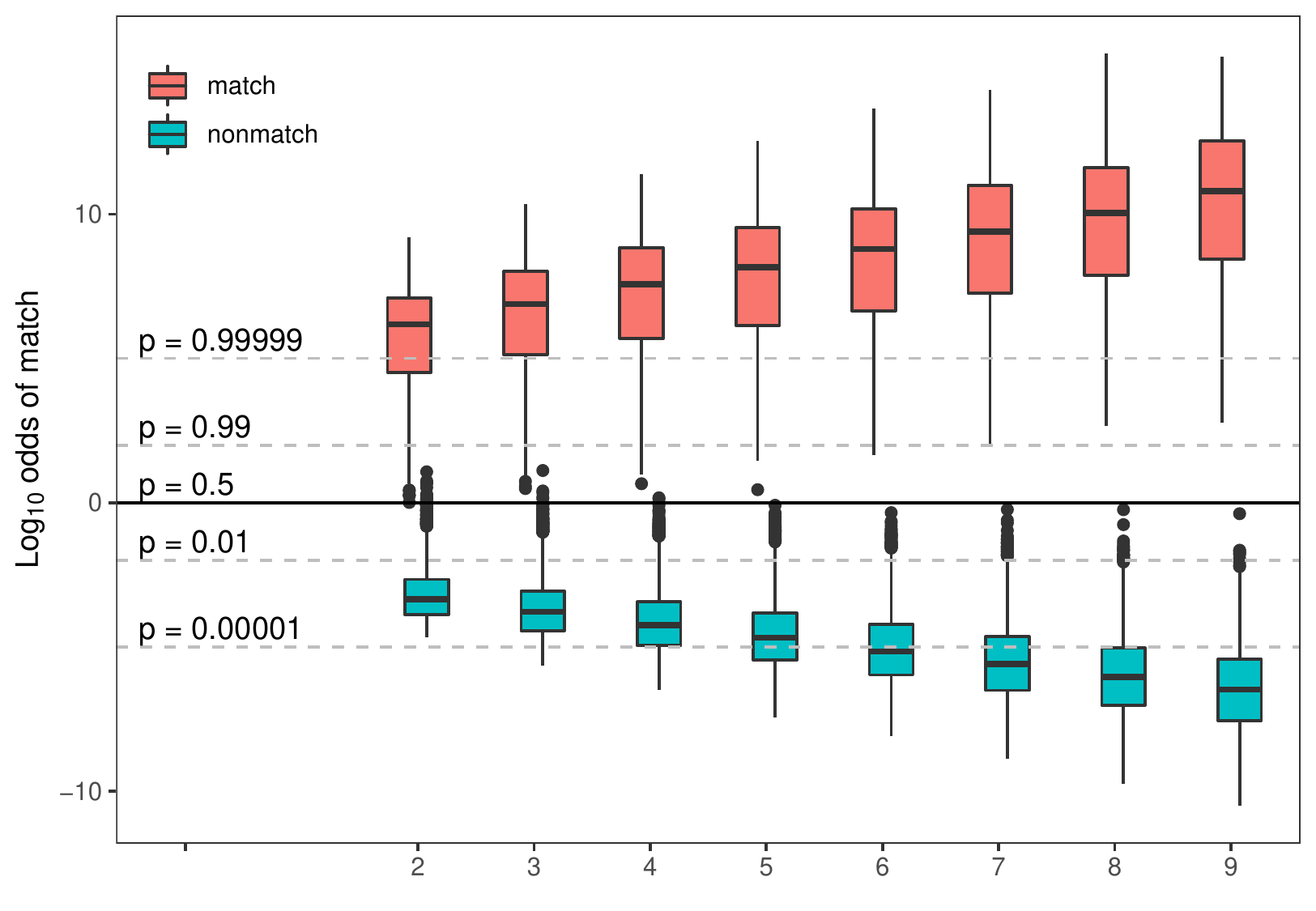}
  \caption{Distributions of the log-odds of a match using models trained
    on the four different sets of surfaces and tested on subsets of $k$ consecutive images
    for $k = 2, 3, \ldots, 9$ for a model with 10 degrees of freedom. \label{fig:allsubsets}}
\end{figure}

\paragraph{Amount of overlap}\label{overlap}
\addcontentsline{toc}{paragraph}{Amount of overlap}

Guided by the results of Figure~\ref{fig:allsubsets}, it is apparent that we need at least 5 to 6 images
for adequate discrimination. We reassessed the imaging procedure to gauge the role of the
image-overlap ratio.  The initial experiment involved imaging surfaces using nine images
with 75\% overlap between images, which provides three observations
for each point on the surface, apart from the edges. However, a similar area can
be imaged using 5 images with 50\% overlap, which produces two observations
of each point on the surface apart from the edges, or using 3 non-overlapping images,
which raises the question of whether anything is gained by having an
additional third image of the same area and, if so, what level of overlap
is optimal.

\begin{figure}[ht]
  \centering
  \includegraphics[width=\textwidth]{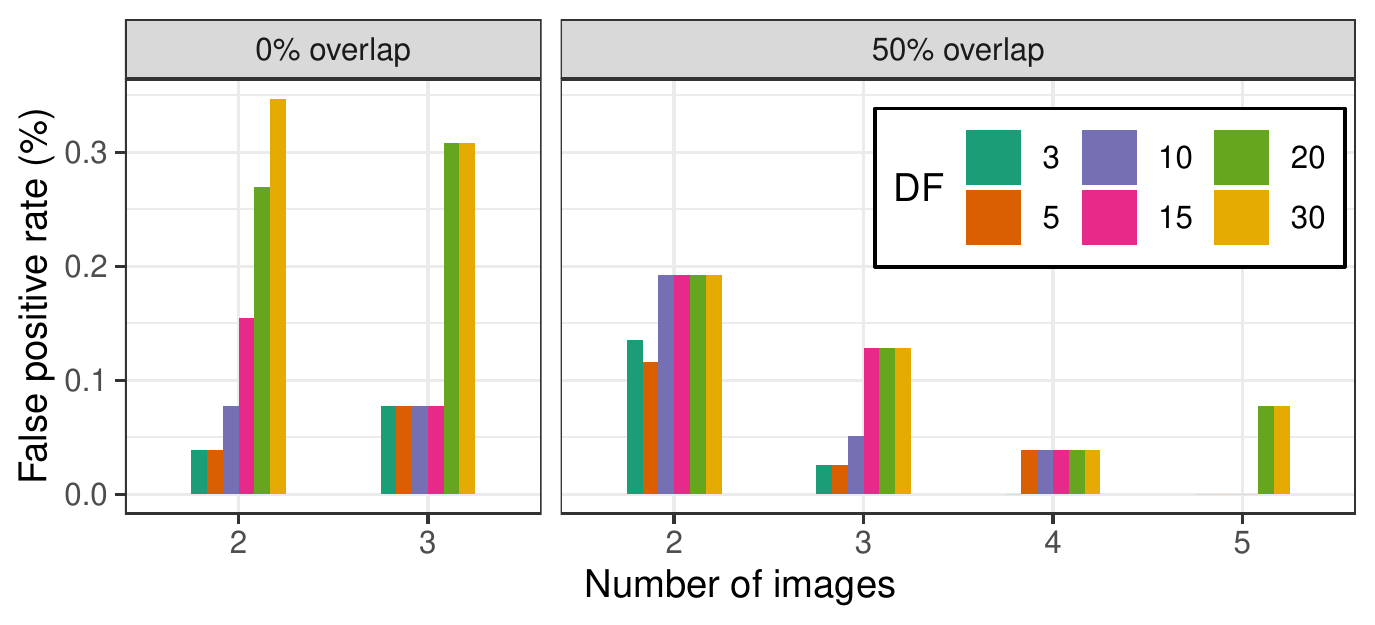}
  \caption{Rates of false positive classifications (in \%) using models trained
    on the four different sets of surfaces using only the images with at most 50\% overlap
    and tested  on subsets of $k$ consecutive images
    for $k = 2, 3, 4,  5$ and using only the 3 non-overlapping images and tested on subsets of $k$ consecutive images for  $k = 2, 3$.\label{fig:fiftyoverlap}}
\end{figure}

We can evaluate this by providing an analysis similar to that done previously:
looking at the classification results when restricted to cases with the specified overlap.
We train classifiers on the same sets as before, except using 5 images with 50\% overlap
instead of 9 images with 75\% overlap and then test the models by classifying
pairs of surfaces using all possible subsets of those images on the surface of sizes 2, 3, 4, and 5.
When restricted to the case of 50\% overlap, there is only perfect classification
when all five images are included and the degrees of freedom ($\nu$) are less than 20 (Figure~\ref{fig:fiftyoverlap}).
 In all cases, there are no false negatives.

We perform a similar exercise in the case of the non-overlapping images. There are
three non-overlapping images per surface which can be used to train classifiers and
the models can then be tested on subsets of those images on each surface of sizes 2 and 3.
In the case of non-overlapping images, none of the models results in perfect classification.
The false positives for each model are also shown in Figure~\ref{fig:fiftyoverlap}.
There are no false negatives in the classification
decisions.

This suggests that, while having more images is generally better,
using 5 images with 50\% overlap appears to be sufficient if all the images are used. Imaging the entire
surface with 50\% overlap outperforms imaging the entire surface
with 75\% overlap in the sense that it works for all of the classes of
model. However, if training with 9 images with 75\% overlap is possible,
testing on new surfaces is feasible with as few as 5 test images with
an appropriate choice of the degrees of freedom parameter in the model.


\subsection*{Conclusions}\label{conclusion}
\addcontentsline{toc}{subsection}{Conclusion}
This paper provides a formal quantitative basis for matching metal
fragments found at crime scenes. Our novel approach 
combines fracture mechanics with statistics and machine learning
to quantify the probability that two candidate specimens are a
match. Our methodology utilizes 3D spectral analysis of the fracture
surface topography, mapped by white light non-contact surface
profilometers. Specifically, our framework realizes the unique attributes of
fracture surfaces at a length scale defined by the fracture process
zone, and uses them to do a quantitative physical match analysis of
metal fragments. Fracture surface morphology has been analyzed for
many classes of materials and shown to be self-affine within a scale relevant to the microscopic scale of the fracture process. We exploit these unique features to quantitatively distinguish the microscopic features on fracture surfaces.   Statistical learning tools are used to classify specimens. 

Using at least 5-6 images in the case of 75\% image overlap or five images with 50\% image overlap, we found that the
matrix-variate t-distribution with 10-15 degrees of freedom, and a first-order autoregressive correlation
structure to describe between-image correlation provides
highly-effective discrimination between matching and
 non-matching surface pairs. Our results show the unique
individuality of a pair of fractured surfaces at wavelengths in the
range of $2-8$ grain diameters ($50-200 \mu m$, or the frequency range of
$5-20 mm^{-1}$ for examined tool-steel). Near-perfect discrimination was achieved, even when the
quality of some of the image pairs deteriorated. Such low-quality images arise from high topological details with a large aspect
ratio that might shadow the surrounding details, and might disturb one
of the frequency bands. Despite these difficulties, our
statistical methods using two frequency bands and an extended number of base-tip
image pairs yielded highly-accurate match decisions.  Among the broad range of training sample sets, this domain of unique
 individuality was found to be persistent and easily identified.   Our
 results provide a method for
 performing matching of fragments with recommendations for model
 parameters, procedures for training models on a similar class of
 materials with the same grain sizes, and procedures for testing new
 samples. Repeated imaging on the same surfaces consistently provided
 similar results. Our framework provided near-perfect matching with
 high confidence and so has the potential to be of significant impact,
 providing the ability to introduce more formality into how forensic
 match comparisons are conducted, through a rigorous mathematical
 framework. Our framework is also general enough to be applied, after
 suitable modifications, to a broad range of fractured materials
 and/or toolmarks, with diverse textures and mechanical properties. In
 doing so, we expect our novel methodology and findings to help
 forensic scientists and practitioners place forensic decision-making
 on a firmer scientific footing. This can help formalize the scientific
 basis for conclusive matching of fragments leading to quantitative
 and more objective forensic decisions.

\bibliographystyle{IEEEtran}
\bibliography{pnas-refs}
\section*{Supplementary Materials}

\subsection{Details on Sample Generation and Imaging}\label{app:sampling}
\addcontentsline{toc}{subsection}{Details on Sample Generation and Imaging}

\begin{figure}[ht]
  \centering
  \includegraphics[width=\textwidth]{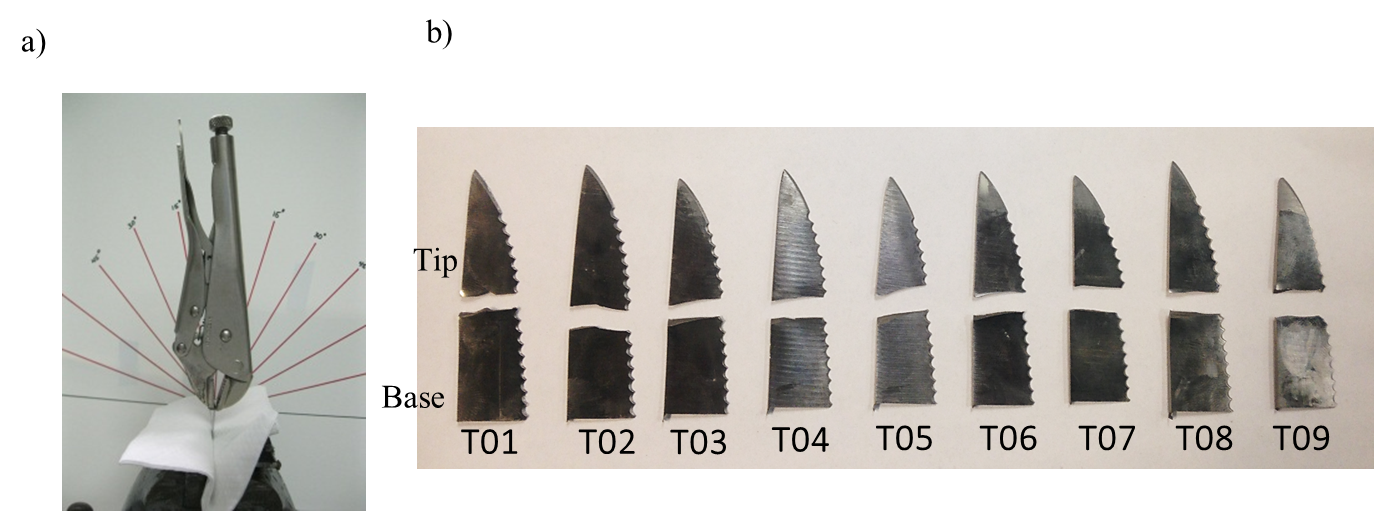}
  \caption{Knife-breaking protocol. (a) Loading fixture. (b) Pairs of knives from the same manufacturer fractured by bending.\label{fig:breaking} }
\end{figure}

\begin{figure}[ht]
  \centering
  \includegraphics[width=\textwidth]{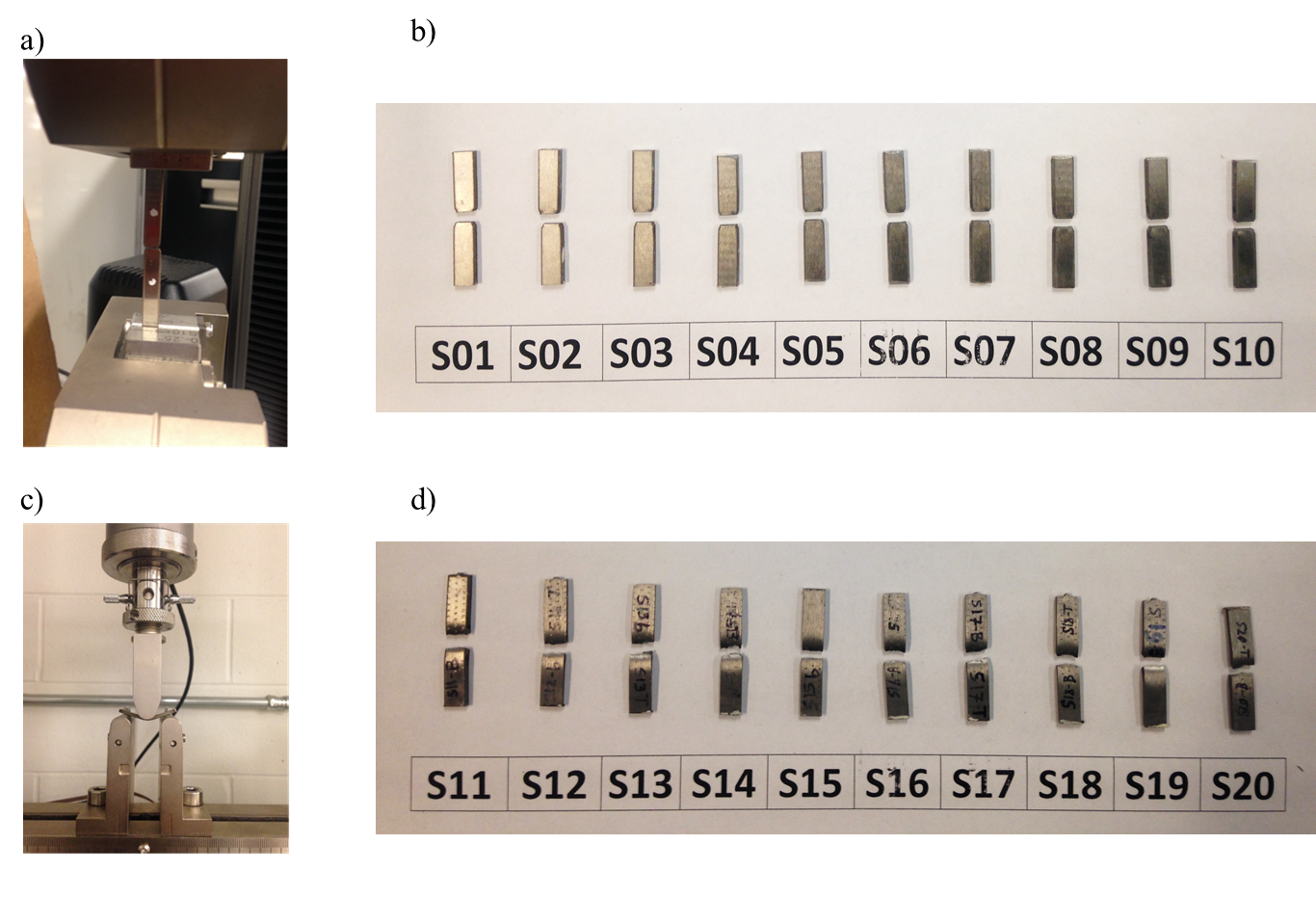}
  \caption{ Sample generation  protocol. (a) sample fracture under a controlled tensile loading, (b) pairs of steel samples fractured by tensile loading, (c) sample fracture under three point bending loading, (d) pairs of steel samples fractured under bending.\label{fig:toolsteel} }
\end{figure}

\begin{figure}[ht]
  \centering
  \includegraphics[width=\textwidth]{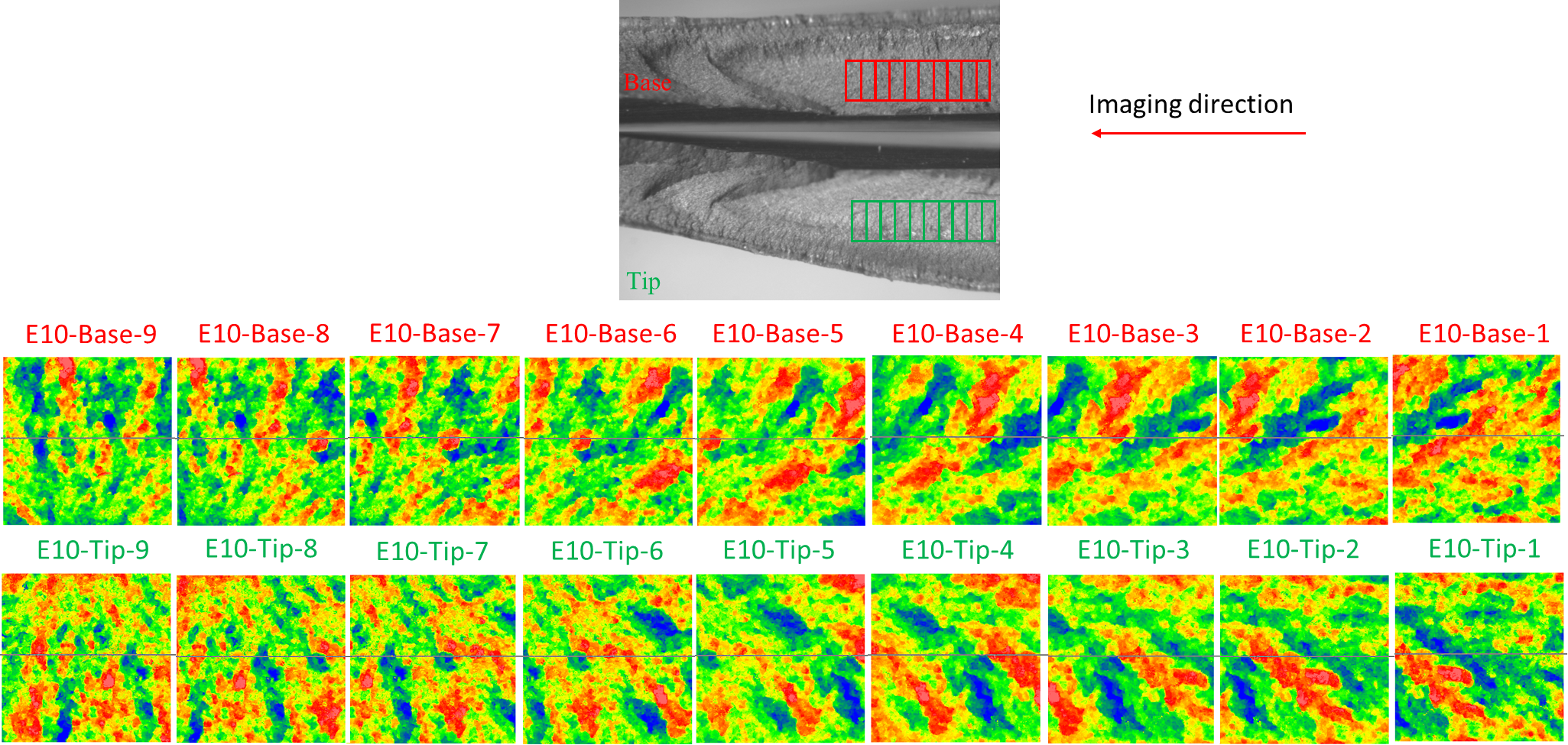}
  \caption{ Nine overlapped and aligned images from a pair of fracture surfaces.\label{fig:imaging} }
\end{figure}

Two main material classes are considered: two sets of nine single serrated
edged knives from the same manufacturer (Chicago Cutlery), and two sets of
ten rectangular (0.25" wide, and 1/16" thick) rods of a common tool steel
material (SS-440C) cut from the same metal sheet to minimize any
variability from the manufacturer. The knives were fractured at random
using a controlled bend fixture shown in Figure~\ref{fig:breaking}(a). a set of fractured
pairs of knives is shown in Figure~\ref{fig:breaking}(b). The two sets of the tool steel
were loaded under either controlled tensile loading at 1 mm/ min
displacement rate (Figure~\ref{fig:toolsteel}(a)) or controlled bending loading at 1.5 $mm/min$ 
displacement rate (Figure~\ref{fig:toolsteel}(c)) until fracture. 
The pairs of tool steel samples fractured by tension and bending are shown
in Figure~\ref{fig:breaking}(b,d), respectively. 
The average grain size for both groups was approximately $dg$ = 25--35 $\mu m$. 
For clarity, we refer to the surface attached to the knife handle as the
base and the surface from the top portion of the knife as the tip. 
The same terminology was applied to the tool steel samples as well. 
The microscopic features of pairs of fracture surfaces were aligned and
analyzed by a standard non-contact 3D optical interferometer (Zygo-NewView
6300), which provides a height resolution of 20 $nm$ and spatial inter-point
resolution of 0.45$\mu m$ (Figure~2(a)). Surface height 3D topographic maps were
acquired from the pairs of fracture surfaces and quantized using spectral
analysis, as summarized in Figure~3.
An extended set of nine topological images with a 550 $\mu  m$ field of view were
collected on each fracture surface, provides a mapping of 0.55 $\mu m/pixel$  is shown in
Figure~\ref{fig:imaging}.
For the collected set of images, digital filters for surface tilt
correction and spike noise removal were applied. 
\FloatBarrier
\FloatBarrier

\subsection{Matrix Variate Normal and Matrix Variate t Distributions}\label{app:definitions}
\addcontentsline{toc}{subsection}{Matrix Normal Definitions}
The matrix variate normal distribution is related to the matrix variate $t$ distribution and is used to
construct it. In this section, we define these distributions as used in the paper.

\begin{definition}
  \label{defn:normal}
   A random $p \times q$ (in this example, $p = 2$ and $q = 9$) matrix $\bX$ has a matrix-variate normal distribution
      with parameters
      $(\bfM,\bSigma,\bOmega$), with $\bfM$ a $p \times q$ matrix specifying the mean,
      $\bSigma$ a $p \times p$ covariance matrix defining the relationship between the rows,
      $\bOmega$ a $q \times q$ covariance matrix defining the relationship between the columns,
  if it has
the probability density function (PDF) 
\begin{align*}f(\bX;\mathbf{M}, \bSigma, \bOmega) &= \frac{\exp\left( -\frac{1}{2} \, \mathrm{tr}\left[ \bOmega^{-1} (\bX - \mathbf{M})^{T} \bSigma^{-1} (\bX - \mathbf{M}) \right] \right)}{(2\pi)^{pq/2} |\bOmega|^{p/2} |\bSigma|^{q/2}}, \end{align*}
where $|\cdot |$ denotes the determinant, $\mathbf{M}$ is a
$p \times q$ matrix that is the mean of $\bX$,
and $\bSigma$ and $\bOmega>0$  describing
the covariances between, respectively, each of the $p$ rows and the
$q$ columns of $\bX$. We write $\bX\sim \mathcal N_{p,q}(\mathbf{M},
\bSigma, \bOmega)$. For identifiability, we set the first element of
$\bSigma$ to be unity.
\end{definition}
The matrix variate normal  distribution can be considered, after rearranging into a
vector (denoted by vec($\bX$)), to be from a multivariate normal (MVN)
distribution with a Kronecker product covariance structure
\citep{gupta1999matrix}. So, if $\bX\sim \mathcal N_{p,q}(\mathbf{M},
\bSigma, \bOmega)$, then $\mathrm{vec}(\bX) \sim
\mN_{pq}(\mathrm{vec}(\mathbf{M}), \bOmega \otimes \bSigma).$

In the case of over-dispersion, that is, higher variance than can be
explained by a normal model, it may be appropriate to use a distribution
with fatter tails, such as a $t$ distribution. A matrix variate $t$ distribution,
here abbreviated as MxV$t$, can be defined as follows:

    \begin{definition}
      \label{defn:t}
      A random $p \times q$ matrix $\bX$ has a MxV$t$ distribution
      with parameters
      $(\bfM,\bSigma,\bOmega$) of similar order as in
      Definition~\ref{defn:normal} (with $\bSigma$ and $\bOmega > 0$) and degrees of freedom (df) $\nu\geq1$  if its PDF is  
\begin{equation}f(\bX;\nu, \mathbf{M}, \bSigma, \bOmega) =
             {\frac{\Gamma _{p}\left({\frac  {\nu +p+q-1}{2}}\right)}{(\pi )^{{\frac  {pq}{2}}}\Gamma _{p}\left({\frac  {\nu +p-1}{2}}\right)}}|
               \bOmega |^{{-{\frac  {p}{2}}}}|\bSigma |^{{-{\frac  {q}{2}}}} 
              \left|{\mathbf  {I}}_{p}+  \bSigma^{{-1}}(\bX-{\mathbf{M}})\bOmega^{{-1}}(\bX-{\mathbf  {M}})^{{{\rm {T}}}}\right|^{{-{\frac  {\nu +p+q-1}{2}}}}.\label{mleqn}\tag{S1}\end{equation}
We use the notation   $\bX\sim
t_{p,q}(\nu,\bfM,\bSigma,\bOmega)$  to indicate that $\bX$ has
this density. 
\end{definition}
         We mention some properties of the MxV$t$
          distribution relevant to this paper.
          \begin{enumerate}
            \item \label{prop1}
              For $p = 1$ and $\bSigma \equiv \nu$ (or $q = 1$ and
              $\bOmega \equiv \nu$), the MxV$t$ distribution reduces to its
              vector-multivariate $t$ (MVT) cousin.
\item \label{prop2} Let the random matrix $\bS\sim  {\mathcal W}_p(\nu + p -1, \bSigma^{-1})$,
  where ${\mathcal W}_p(\kappa,\bPsi)$ is the $p\times p$-dimensional
  Wishart distribution with d.f. $\kappa$ and scale matrix $\bPsi$. If
  $\bX\mid \bS\sim {\mathcal N}_{p,q}(\bfM,\bS^{-1}, \bOmega)$,
  then  $\bX\sim t_{p,q}(\nu, \bfM, \bSigma, \bOmega)$~(see
  \cite{gupta1999matrix}, page 135).
Further, $\bS\mid\bX \sim {\mathcal W}_p(\nu+p+q-1, [(\bX-\mathbf{M})\bOmega^{-1}(\bX-\mathbf{M})^T + \bSigma]^{-1})$~\citep{iranmanesh2010conditional}.
\end{enumerate}

             As $\nu \to \infty$, the MxV$t$ converges to a matrix variate normal distribution.
             In this application, the correlations between individual images are taken to be
             identically distributed within each class (true matches and true non-matches), which implies for
             each class the $\bfM$ matrix is constant along its rows and that the $\bOmega$ matrix
             can be expressed as a correlation matrix with a unit diagonal. We further assume that the
             covariance of the observations between neighbors remain the same across the imaged surface
             (e.g., the relationship between the first and second images is the same as the relationship
             between the eighth and ninth images). Because of the overlapping structure, we use
             an autoregressive covariance structure to describe the covariance and use an AR(1)
             model in this application. In an AR(1) correlation matrix, the correlation of any two
             adjacent elements is $\rho$ and for two elements, $a_i$ and $a_j$, the correlation between
             them is $\rho^{|i - j|}$, with $|\rho|< 1$.

\subsection{Tables of Results}\label{app:tables}
\addcontentsline{toc}{subsection}{Tables of Results}

Each table contains classification results for different settings
of the $\nu$ parameter and $k$ parameter.
Four different data sets were used as training sets: two sets with 9 samples and 9 images per sample (with 75\% overlap)
and two sets with 10 samples and 9 images per sample (with 75\% overlap). The sets with 9 samples, then,
had 81 sets of comparisons between base and tip: 9 true matches and 72 true non-matches.
The sets with 10 samples
had 100 sets of comparisons between base and tip: 10 true matches and 90 true non-matches.
In the first table, all 9 images (with 75\% overlap) for every surface are included in the model training. 
In the second table, only 5 images with 50\% overlap are included in the model training.
In the final table, only 3 non-overlapping images are included.
For each setting of $\nu$ and for each of the four different datasets, 
we train a model and then test on all four datasets by classifying all consecutive subsets of images of size $k$.
In the tables, the first two columns indicate the value of $\nu$ and the number of consecutive images, $k$, used to classify test samples. The final two columns tally how many were truly matching or truly non-matching comparisons. 
All models used equal priors and a classification threshold of 0 log-odds (equivalent to a posterior probability of 0.5).

To compute the False Positive Rate (FPR) for any row, divide the number in the False Pos column by the number of True Non-Matches
(equivalently, by the sum of False Pos and True Neg columns). Other statistics can be computed in a similar manner for each row.
\begin{table*}
  \caption{Summary of match and non-match decisions for the models fitted using all 9 images and tested on
all consecutive subsets of images of size $k$.}
\begin{tabular}{lrrrrrrr}

Model & k & False Pos & False Neg & True Pos & True Neg & True Match & True Non-Match\\
\hline
$\nu$ = 3 & 2 & 12 & 5 & 1211 & 10356 & 1216 & 10368 \\
$\nu$ = 3 & 3 & 3 & 4 & 1060 & 9069 & 1064 & 9072 \\
$\nu$ = 3 & 4 & 3 & 1 & 911 & 7773 & 912 & 7776 \\
$\nu$ = 3 & 5 & 0 & 3 & 757 & 6480 & 760 & 6480 \\
$\nu$ = 3 & 6 & 0 & 0 & 608 & 5184 & 608 & 5184 \\
$\nu$ = 3 & 7 & 0 & 0 & 456 & 3888 & 456 & 3888 \\
$\nu$ = 3 & 8 & 0 & 0 & 304 & 2592 & 304 & 2592 \\
$\nu$ = 3 & 9 & 0 & 0 & 152 & 1296 & 152 & 1296 \\
$\nu$ = 5 & 2 & 13 & 4 & 1212 & 10355 & 1216 & 10368 \\
$\nu$ = 5 & 3 & 3 & 0 & 1064 & 9069 & 1064 & 9072 \\
$\nu$ = 5 & 4 & 3 & 0 & 912 & 7773 & 912 & 7776 \\
$\nu$ = 5 & 5 & 0 & 1 & 759 & 6480 & 760 & 6480 \\
$\nu$ = 5 & 6 & 0 & 0 & 608 & 5184 & 608 & 5184 \\
$\nu$ = 5 & 7 & 0 & 0 & 456 & 3888 & 456 & 3888 \\
$\nu$ = 5 & 8 & 0 & 0 & 304 & 2592 & 304 & 2592 \\
$\nu$ = 5 & 9 & 0 & 0 & 152 & 1296 & 152 & 1296 \\
$\nu$ = 10 & 2 & 14 & 0 & 1216 & 10354 & 1216 & 10368 \\
$\nu$ = 10 & 3 & 5 & 0 & 1064 & 9067 & 1064 & 9072 \\
$\nu$ = 10 & 4 & 3 & 0 & 912 & 7773 & 912 & 7776 \\
$\nu$ = 10 & 5 & 0 & 0 & 760 & 6480 & 760 & 6480 \\
$\nu$ = 10 & 6 & 0 & 0 & 608 & 5184 & 608 & 5184 \\
$\nu$ = 10 & 7 & 0 & 0 & 456 & 3888 & 456 & 3888 \\
$\nu$ = 10 & 8 & 0 & 0 & 304 & 2592 & 304 & 2592 \\
$\nu$ = 10 & 9 & 0 & 0 & 152 & 1296 & 152 & 1296 \\
$\nu$ = 15 & 2 & 18 & 0 & 1216 & 10350 & 1216 & 10368 \\
$\nu$ = 15 & 3 & 7 & 0 & 1064 & 9065 & 1064 & 9072 \\
$\nu$ = 15 & 4 & 4 & 0 & 912 & 7772 & 912 & 7776 \\
$\nu$ = 15 & 5 & 1 & 0 & 760 & 6479 & 760 & 6480 \\
$\nu$ = 15 & 6 & 0 & 0 & 608 & 5184 & 608 & 5184 \\
$\nu$ = 15 & 7 & 0 & 0 & 456 & 3888 & 456 & 3888 \\
$\nu$ = 15 & 8 & 0 & 0 & 304 & 2592 & 304 & 2592 \\
$\nu$ = 15 & 9 & 0 & 0 & 152 & 1296 & 152 & 1296 \\
$\nu$ = 20 & 2 & 19 & 0 & 1216 & 10349 & 1216 & 10368 \\
$\nu$ = 20 & 3 & 9 & 0 & 1064 & 9063 & 1064 & 9072 \\
$\nu$ = 20 & 4 & 4 & 0 & 912 & 7772 & 912 & 7776 \\
$\nu$ = 20 & 5 & 1 & 0 & 760 & 6479 & 760 & 6480 \\
$\nu$ = 20 & 6 & 0 & 0 & 608 & 5184 & 608 & 5184 \\
$\nu$ = 20 & 7 & 1 & 0 & 456 & 3887 & 456 & 3888 \\
$\nu$ = 20 & 8 & 1 & 0 & 304 & 2591 & 304 & 2592 \\
$\nu$ = 20 & 9 & 1 & 0 & 152 & 1295 & 152 & 1296 \\
$\nu$ = 30 & 2 & 20 & 0 & 1216 & 10348 & 1216 & 10368 \\
$\nu$ = 30 & 3 & 9 & 0 & 1064 & 9063 & 1064 & 9072 \\
$\nu$ = 30 & 4 & 4 & 0 & 912 & 7772 & 912 & 7776 \\
$\nu$ = 30 & 5 & 1 & 0 & 760 & 6479 & 760 & 6480 \\
$\nu$ = 30 & 6 & 0 & 0 & 608 & 5184 & 608 & 5184 \\
$\nu$ = 30 & 7 & 1 & 0 & 456 & 3887 & 456 & 3888 \\
$\nu$ = 30 & 8 & 1 & 0 & 304 & 2591 & 304 & 2592 \\
$\nu$ = 30 & 9 & 1 & 0 & 152 & 1295 & 152 & 1296 \\
\end{tabular}

\label{tab:nineimages}
\end{table*}
\begin{table}
  \caption{Summary of match and non-match decisions for the models fitted using 5 images with at most 50\% overlap and tested on
all consecutive subsets of images of size $k$.}
\begin{tabular}{lrrrrrrr}
Model & k & False Pos & False Neg & True Pos & True Neg & True Match & True Non-Match\\
\hline

$\nu$ = 3 & 2 & 7 & 0 & 608 & 5177 & 608 & 5184 \\
$\nu$ = 3 & 3 & 1 & 0 & 456 & 3887 & 456 & 3888 \\
$\nu$ = 3 & 4 & 0 & 0 & 304 & 2592 & 304 & 2592 \\
$\nu$ = 3 & 5 & 0 & 0 & 152 & 1296 & 152 & 1296 \\
$\nu$ = 5 & 2 & 6 & 0 & 608 & 5178 & 608 & 5184 \\
$\nu$ = 5 & 3 & 1 & 0 & 456 & 3887 & 456 & 3888 \\
$\nu$ = 5 & 4 & 1 & 0 & 304 & 2591 & 304 & 2592 \\
$\nu$ = 5 & 5 & 0 & 0 & 152 & 1296 & 152 & 1296 \\
$\nu$ = 10 & 2 & 10 & 0 & 608 & 5174 & 608 & 5184 \\
$\nu$ = 10 & 3 & 2 & 0 & 456 & 3886 & 456 & 3888 \\
$\nu$ = 10 & 4 & 1 & 0 & 304 & 2591 & 304 & 2592 \\
$\nu$ = 10 & 5 & 0 & 0 & 152 & 1296 & 152 & 1296 \\
$\nu$ = 15 & 2 & 10 & 0 & 608 & 5174 & 608 & 5184 \\
$\nu$ = 15 & 3 & 5 & 0 & 456 & 3883 & 456 & 3888 \\
$\nu$ = 15 & 4 & 1 & 0 & 304 & 2591 & 304 & 2592 \\
$\nu$ = 15 & 5 & 0 & 0 & 152 & 1296 & 152 & 1296 \\
$\nu$ = 20 & 2 & 10 & 0 & 608 & 5174 & 608 & 5184 \\
$\nu$ = 20 & 3 & 5 & 0 & 456 & 3883 & 456 & 3888 \\
$\nu$ = 20 & 4 & 1 & 0 & 304 & 2591 & 304 & 2592 \\
$\nu$ = 20 & 5 & 1 & 0 & 152 & 1295 & 152 & 1296 \\
$\nu$ = 30 & 2 & 10 & 0 & 608 & 5174 & 608 & 5184 \\
$\nu$ = 30 & 3 & 5 & 0 & 456 & 3883 & 456 & 3888 \\
$\nu$ = 30 & 4 & 1 & 0 & 304 & 2591 & 304 & 2592 \\
$\nu$ = 30 & 5 & 1 & 0 & 152 & 1295 & 152 & 1296 \\

\end{tabular}

\label{tab:fiveimages}
\end{table}
\begin{table}
   \caption{Summary of match and non-match decisions for the models fitted using 3 adjacent non-overlapping images and tested on
all consecutive subsets of images of size $k$.}
     \begin{tabular}{lrrrrrrr}
       Model & k & False Pos & False Neg & True Pos & True Neg & True Match & True Non-Match\\
       \hline
$\nu$ = 3 & 2 & 1 & 0 & 304 & 2591 & 304 & 2592 \\
$\nu$ = 3 & 3 & 1 & 0 & 152 & 1295 & 152 & 1296 \\
$\nu$ = 5 & 2 & 1 & 0 & 304 & 2591 & 304 & 2592 \\
$\nu$ = 5 & 3 & 1 & 0 & 152 & 1295 & 152 & 1296 \\
$\nu$ = 10 & 2 & 2 & 0 & 304 & 2590 & 304 & 2592 \\
$\nu$ = 10 & 3 & 1 & 0 & 152 & 1295 & 152 & 1296 \\
$\nu$ = 15 & 2 & 4 & 0 & 304 & 2588 & 304 & 2592 \\
$\nu$ = 15 & 3 & 1 & 0 & 152 & 1295 & 152 & 1296 \\
$\nu$ = 20 & 2 & 7 & 0 & 304 & 2585 & 304 & 2592 \\
 $\nu$ = 20 & 3 & 4 & 0 & 152 & 1292 & 152 & 1296 \\
$\nu$ = 30 & 2 & 9 & 0 & 304 & 2583 & 304 & 2592 \\
$\nu$ = 30 & 3 & 4 & 0 & 152 & 1292 & 152 & 1296 \\
     \end{tabular}

 \label{tab:threeimages}
 \end{table}


\end{document}